\documentclass[final]{svjour2}
\usepackage{graphicx}
\usepackage{bm}
\usepackage{rotating}
\usepackage{amssymb}
\usepackage{mathptmx}
\usepackage[numbers]{natbib}
\makeatletter
\journalname{Journal of Low Temperature Physics}

\bibpunct{}{}{,}{s}{}{,}

\begin{document}

\newcommand{\hdblarrow}{H\makebox[0.9ex][l]{$\downdownarrows$}-}
\newcommand{\eqnum}{\addtocounter{equation}{1}\tag*{\normalsize{(\arabic{equation})}}}
\title{Expansion of a Bose-Einstein Condensate with Vortices}

\author{R.Tsuchitani \and M. Tsubota}

\institute{
Department of Physics, Osaka City University,\\ Sumiyoshi-ku, Osaka 558-8585, Japan\\
Tel.: +81-6-6605-3073\\ Fax: +81-6-6605-2522\\
\email{tsubota@sci.osaka-cu.ac.jp}\\
}

\date{08.11.2013}

\maketitle

\keywords{Bose-Einstein Condensates, Vortices}

\begin{abstract}
The expansion of Bose-Einstein condensates (BECs) is numerically studied. Usually, the aspect ratio of a condensate is inverted due to the anisotropy of the uncertainty principle. In turbulent BECs, however, the aspect ratio is observed to remain constant. The reason for this phenomenon is still unknown, being a challenging topic in the study of quantum turbulence. Here, the two-dimensional Gross-Pitaevskii equation is solved and the time development of the aspect ratio and of the radii of the condensates are calculated in the presence of vortices. The results indicate that vortex pairs must be added to a condensate in order to expand it with keeping its aspect ratio. The anisotropy due to the vortex pairs competes with that due to the uncertainty principle so that the aspect ratio remains constant. 
\end{abstract}

\section{Introduction}
Quantum turbulence (QT) has been studied in superfluid $^{4}$He and $^{3}$He for many years. It is composed of topological defects known as quantized vortices. Such vortices have been investigated in superfluid $^{4}$He both experimentally and theoretically.\cite{dipole} However, it is difficult to control the turbulent state in helium and determine the vortex configuration. An alternative system in which quantized vortices have been studied is magnetically or optically trapped atomic BECs.\cite{gp} A BEC is weakly interacting, making it easier to treat theoretically. Furthermore, many physical parameters of BECs are experimentally controllable and various physical quantities such as the density and phase of a BEC can be directly observed. For these reasons, QT in atomic BECs has been studied.\cite{experiment,analysis,kobayashi,Berloff,laser}

Using the method similar to those of Kobayashi {\it et al.},\cite{kobayashi} Henn {\it et al.} generated QT experimentally\cite{experiment}. They created a condensate in an axisymmetric harmonic trap and generated turbulence by vibrating and rotating it. Then they turned off the trap and let the condensate freely expand. In ordinary BECs, the aspect ratio of the condensate inverts with expanding, because a narrow spatial distribution implies a broad momentum distribution according to the Heisenberg uncertainty principle.\cite{gp} In turbulent BECs, however, the aspect ratio of the condensate surprisingly remained constant. Henn {\it et al.} relied on the constant aspect ratio to determine whether turbulence was being created, here referred to as "self-similar expansion (SSE)".

Caracanh$\tilde{a}$s {\it et al.} theoretically analyzed SSE\cite{analysis}. They derived the vortex contribution to the hydrodynamic equations from a Lagrangian in the Gross-Pitaevskii (GP) model, by considering the additional energy of an entangled vortex configuration. They assumed a trial form for the macroscopic wavefunction $\psi$ of the GP equation\cite{gp} that depends on the variational parameters (specifically, the time-dependent condensate radii and the expansion variables) to derive an equation of motion for the radii of the condensates. It reproduced the SSE of BECs.

The present numerical study of SSE introduces two innovations. Firstly, the two-dimensional GP equation is solved directly. As a consequence, the time development of the condensates is realistically calculated as long as the system is almost two-dimensional. Secondly, we can prepare vortices in arbitrary positions and investigate how the SSE depends on their configurations.

The directivity of our work is the following. We study the SSE of two-dimensional BECs with a few number of vortices, not addressing a turbulent BEC. However, the size of trapped BECs is generally not much larger than the core size of vortices. Hence, even a system with a few number of vortices yields relatively high number density of vortices. Therefore, even the simulation of a BEC with a few number of vortices should give some useful insight for the SSE.

The present analysis is based on the experiments of Henn {\it et al.}\cite{experiment}  We calculate the dynamics of the aspect ratio of a condensate with vortices. Another kind of anisotropy is needed to reduce the anisotropy due to the uncertainty principle. We propose a simple model by adding two vortex pairs to the condensates to reproduce the SSE. 
\section{Formulation}
A BEC at zero temperature can be accurately described by the GP equation\cite{gp} with a macroscopic wavefunction $\psi$. For simplicity, we restrict attention here to the two-dimensional GP equation,
\begin{eqnarray}
i\hbar\frac{\partial}{\partial t}\psi = -\frac{\hbar^2}{2M}\nabla^2\psi + V\psi + g_{2D}|\psi|^2\psi,
\label{eq:gp_2d}
\end{eqnarray}  
where $M$, $V$, and $g_{2D}$ are the mass of a particle, the harmonic potential, and the interaction coefficient, respectively. The harmonic potential is
\begin{eqnarray}
V &=& \frac{1}{2}M({\omega_x}^2x^2+{\omega_y}^2y^2) = \frac{1}{2}M\omega^2(x^2+\lambda^2y^2),
\end{eqnarray}
where $\lambda=\omega_y/\omega_x$ is the anisotropic parameter. 
Consider experiments\cite{experiment} using $^{87}$Rb, for which $M=1.44\times10^{-25}$ kg, $N = 10^5$ is the number of particles, and $\omega = 2\pi\times210$ rad/s.

We define the aspect ratio in terms of the dispersion of the density distribution ${\sigma_i}^2$ $(i=x,y)$ along the $x$ and $y$ axes as,
\begin{eqnarray}
{\sigma_i}^2 = \int_{-\infty}^{\infty} \left(r_i - \bar{r}_i \right)^2 |\psi|^2 dr_i,
\end{eqnarray}
where $\bar{r}_i = \int_{-\infty}^{\infty} r_i |\psi|^2 dr_i$ is the mean value.
The radius $R_i$ of the condensate is taken to be 
$R_i = 3\sigma_i$ with a standard deviation $\sigma_i$.
Then the aspect ratio is defined as 
$R_x/R_y = \sigma_x/\sigma_y$.

Using the Crank-Nicholson method, we numerically solve Eq. (\ref{eq:gp_2d}). 
The coordinate is normalized by the length $a_h=\left(\hbar/M\omega\right)^{1/2} = 0.745 \mu$m, where the size of the box is $102.4a_h\times102.4a_h$. The spatial grid in the $x$ and $y$ directions is discretized into $2048\times2048$ bins. The time is normalized by the frequency $\omega$ of the harmonic potential in the $xy$ plane.
\section{Numerical Calculations}
\begin{figure}[h!]
\centering
\includegraphics[width=110mm,height=50mm]{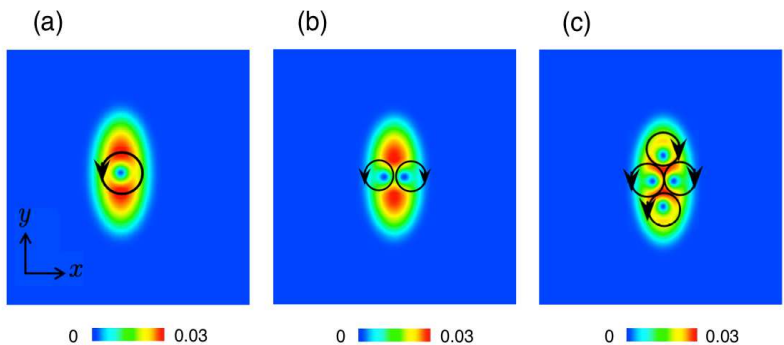}
\caption{(Color online) Density profile of condensates in their ground states at $t=0$. (a) Condensate with a single vortex. (b) Condensate with a vortex pair along the $x$ axis. (c) Condensate with two vortex pairs along the $x$ and $y$ axes. The black arrows on the circles show the direction of circulation of each vortex. The anisotropic parameter $\lambda$ is $0.5$. The scale of each panel is same and the normalized Thomas-Fermi radii $R_x$ and $R_y$ of the condensates are respectively about 5.0 and 9.0,  although they change somewhat by the vortex configuration.}
\label{fig:fig1}
\end{figure}

Condensates are prepared in the ground states trapped by the harmonic potential $V$. After turning off the harmonic potential at $t=0$, the condensates freely expand and the time development of the aspect ratio is tracked. An important issue is how the expansion depends on the vortex configuration. If necessary, we prepare vortices in the initial state by a method; we diminish the density at a point where we place a vortex and then paste the  circular phase on the point. 

First, a condensate is prepared in the absence of vortices, and the inversion of the aspect ratio is confirmed.
Second, a single vortex is added to the center of the condensate as shown in Fig. \ref{fig:fig1}(a), and the effects of the centrifugal force are monitored due to superfluid flow around the vortex.
Third, one or two vortex pairs are added like Fig.1 (b) and (c) to observe the suppression of the inversion of the aspect ratio. The time-development of the aspect ratio comes from the competition between the effects of anisotropy due to the uncertainty principle and the vortex pairs. To extract the anisotropy due to vortex pairs, a circular condensate is considered here, which lowers the effect of the anisotropy due to the uncertainty principle.

\begin{figure}[h!]
\centering
\includegraphics[width=110mm]{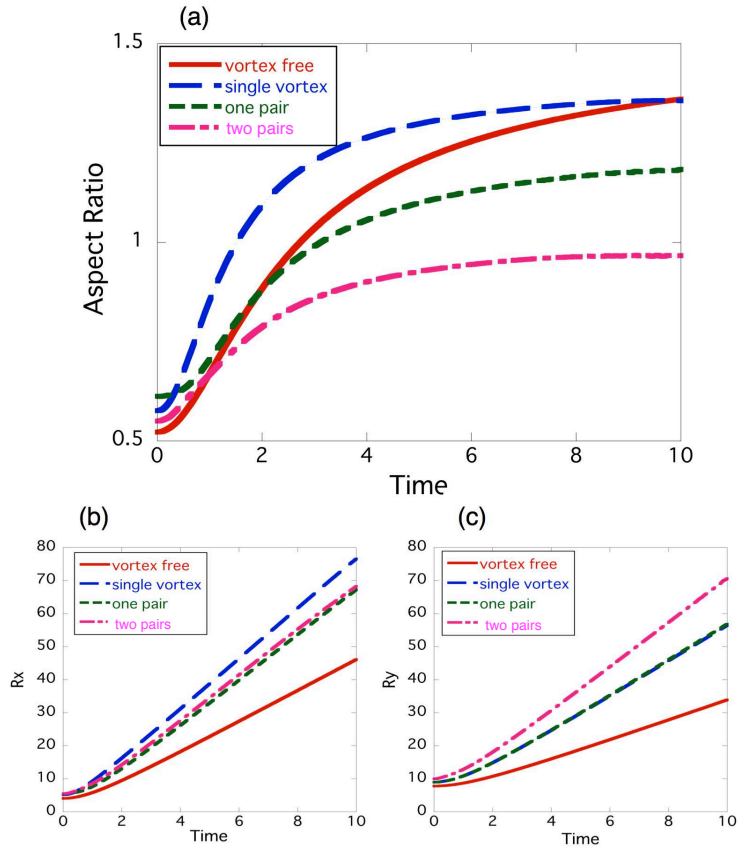}
\caption{(Color online) Time development of the condensates after turning off the harmonic trap at $t=0$. Time development of (a) the aspect ratio, and (b) and (c) the radii. In (c) the curve of "one pair" overlaps that of "single vortex".}
\label{fig:fig2}
\end{figure}

To confirm that this two-dimensional calculation is relevant to the actual three-dimensional phenomena, we address a vortex-free condensate in the first place. Figure \ref{fig:fig2}(a) demonstrates that the aspect ratio inverts during the free expansion because of the anisotropy from the uncertainty principle. The time at which the aspect ratio exceeds unity is $2.1$ ms, comparable to the $7.6$ ms observed experimentally.\cite{experiment}

The aspect ratio of a single-vortex condensate is found to invert faster than does a vortex-free condensate. Figures \ref{fig:fig2}(b) and (c) graph the time development of the radii $R_x$ and $R_y$. A single-vortex condensate expands faster than a vortex-free one because the centrifugal force from the superfluid flow around the vortex accelerates the expansion. The circular superfluid  cannot reduce the anisotropy due to the uncertainty principle. The repulsive interaction as well as the kinetic energy contributes to the time development of all of the aspect ratio, the $R_x$ and the $R_y$. In both cases of the absence and presence of vortices, the increase of the repulsive interaction makes the speed of the expansion faster\cite{analysis}, while it does not generate any anisotropy to reduce the one due to the uncertainty principle because the interaction is isotropic. Hence, the interaction cannot be the main mechanism causing SSE.

Adding vortex pairs to condensates brings in another kind of anisotropy that competes with that due to the uncertainty principle. To investigate this, the one-pair condensate in Fig. \ref{fig:fig1}(b) and the two-pairs condensate in panel (c) are prepared. The two-pairs condensate can give some clues to turbulent BECs because it is well known that the initial stage of turbulence can appear from quadrupole BEC mode decay and the two-pairs condensate can be regarded as a quadrupole BEC. Figure \ref{fig:fig2}(a) shows that the inversion of the aspect ratio is increasingly suppressed as the number of vortex pairs is increased. The aspect ratio of two-pair condensate in Fig. \ref{fig:fig2}(a) converges to unity because as shown in Fig. \ref{fig:fig1}(c) one vortex pair ($x$-axis) is put perpendicular to the other pair ($y$-axis) and the resulting velocity field $v_x$ and $v_y$ generated by the two vortex pairs is approximately equal. We do not know properly the reason why the aspect ratio goes to unity eventually. This is probably because the effect of the two vortex pairs may become superior to that of the uncertainty principle. 

\begin{figure}[h!]
\centering
\includegraphics[width=120mm,height=65mm]{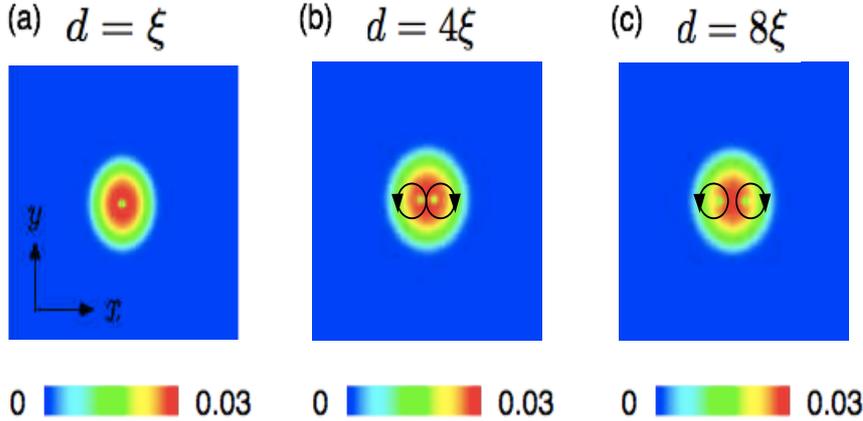}
\caption{(Color online) Density profiles of the condensates with a dipole of the distance (a) $d=\xi$, (b) $d=4\xi$ and (c) $d=8\xi$. In (a), the two vortices are close to overlap. The black arrows on the circles show the direction of circulation of each vortex.}
\label{fig:fig10}
\end{figure}

\begin{figure}[h!]
\centering
\includegraphics[width=110mm]{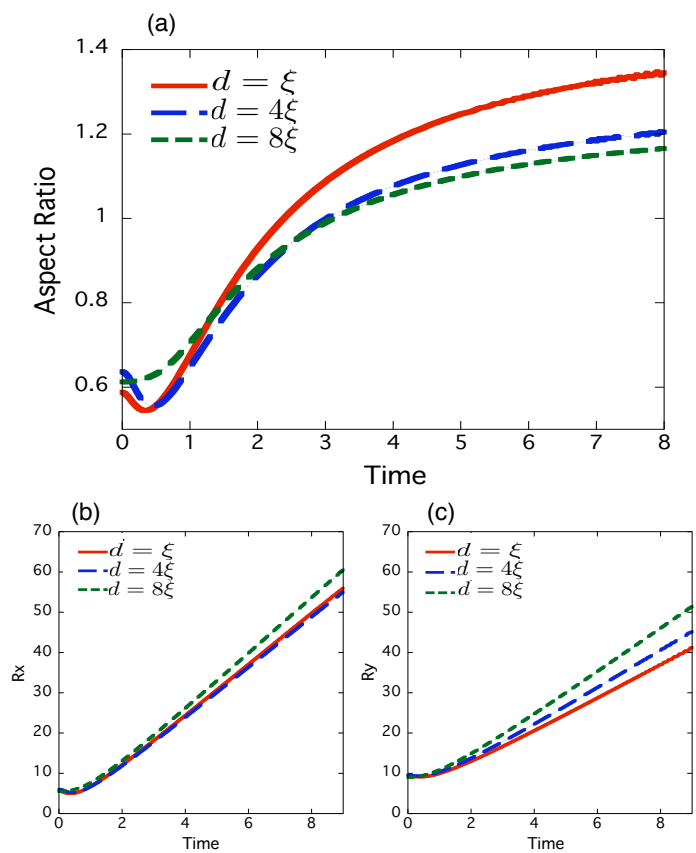}
\caption{(Color online) Dependence on the distance $d$ between the vortices in a pair. Time development of (a) the aspect ratio, and (b) and (c) the radii $R_x$ and $R_y$.}
\label{fig:fig3}
\end{figure}

A vortex pair can be characterized by the distance $d$, and the dependence of the dynamics on $d$ is studied for three types of pairs shown in Fig. \ref{fig:fig10}. A vortex pair oriented along the $x$ axis results in a faster expansion of the condensate along the $y$ axis than along the $x$ axis, because the vortex pair creates a dipolar velocity field directed along the $y$ axis. The pairs are studied having three different distances ($d=\xi$, $4\xi$, $8\xi$) relative to the coherence length $\xi = \hbar/\sqrt{2m n_0 U_0}=0.25$, where $n_0$ and $U_0$ are the density at the center of a condensate and an interaction coefficient respectively.  
Figure \ref{fig:fig3}(a) demonstrates the increasing suppression of the aspect ratio with the increase in the value of $d$. As shown in Figs. \ref{fig:fig3}(b) and (c), $R_y$ has a larger variation among the three types of pairs than does $R_x$. 

This dependence on $d$ may be attributable to the following. The kinetic energy $K$ of a vortex pair per unit length\cite{dipole}is given by
\begin{eqnarray}
K = \frac{\rho\kappa^2}{2\pi}\ln \left(\frac{d}{a}\right),
\label{eq:kappa}
\end{eqnarray}
where $\rho$, $\kappa$, and $a$ are the density of the condensate, and the circulation and size of the vortex core, respectively. Equation (\ref{eq:kappa}) shows that the kinetic energy increases with $d$, which leads to the differences in $R_y$ between the three types of pairs. Consequently, the aspect ratio $R_x/R_y$ is increasingly suppressed with increasing $d$. In the following calculations, the distance of the vortex pair is fixed at $d=4\xi$.

To quantify the anisotropy due to a vortex pair, a circular condensate is prepared having a pair of vortices along the $x$ axis. 
Such an arrangement enables us to extract only the anisotropy due to vortex pairs.

\begin{figure}[h!]
\centering
\includegraphics[width=120mm]{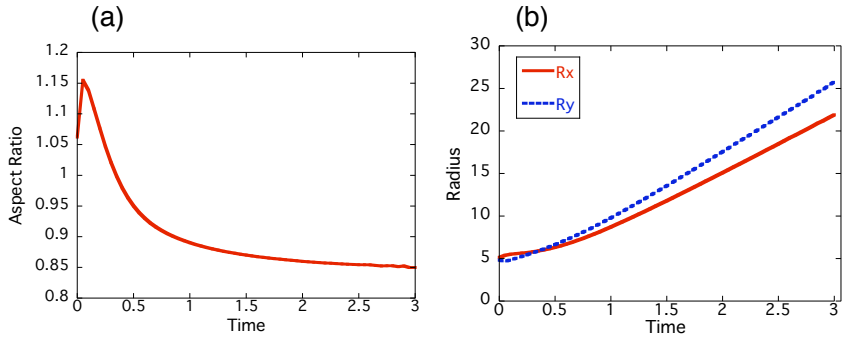}
\caption{(Color online) Time development of (a)  the aspect ratio of a circular condensate having a vortex pair along the $x$ axis, and (b) the radii. The anisotropic parameter $\lambda$ has been increased to $1.0$.}
\label{fig:fig5}
\end{figure}

Figure \ref{fig:fig5}(a) shows that the aspect ratio $R_x/R_y$ decreases overall. This decrease occurs because $R_y$ increases faster than $R_x$, as shown in Fig. \ref{fig:fig5}(b). The initial small raise of $R_x/R_y$ comes from the small decrease in $R_y$, which is an artifact of a vortex pair. When the anisotropy due to the uncertainty principle becomes comparable to that due to the vortex pairs, the aspect ratio of a condensate becomes almost constant.

\begin{figure}[h!]
\centering
\includegraphics[width=120mm]{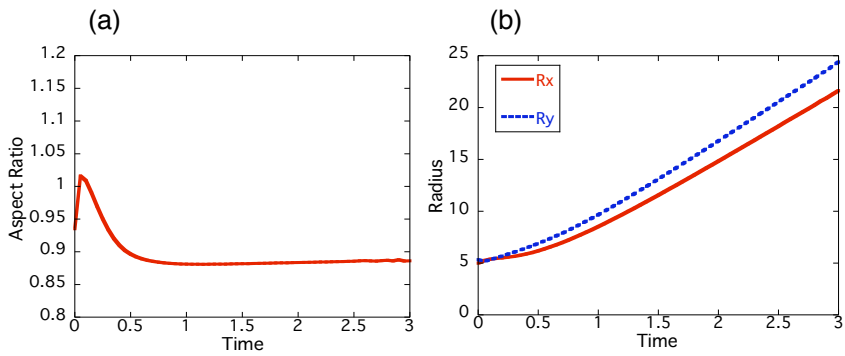}
\caption{(Color online) SSE at $\lambda=0.87$, in which the anisotropy due to the uncertainty principle competes with that due to the vortex pair. The aspect ratio quickly levels off in value. Time development of (a) the aspect ratio, and (b) the radii $R_x$ and $R_y$.}
\label{fig:fig6}
\end{figure}

\begin{figure}[h!]
\centering
\includegraphics[width=95mm]{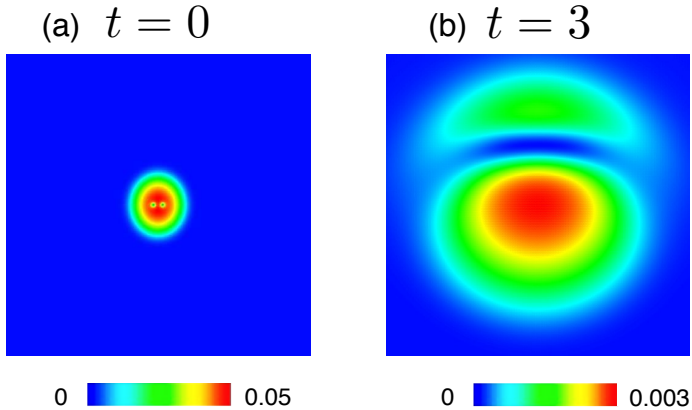}
\caption{(Color online) Density profiles of a one-pair condensate at $\lambda=0.87$. (a) Density profile at $t=0$. (b) Density profile at $t=3$.}
\label{fig:fig7}
\end{figure}

Finally, we fix the initial position of a vortex pair and its distance $d=4\xi$, and consider what happens if we change the value of $\lambda$.
As indicated in Fig. \ref{fig:fig6}(a), the aspect ratio levels off at $\lambda=0.87$. This constancy arises because the anisotropy due to the vortex pair competes with that due to the uncertainty principle. Figure \ref{fig:fig7} shows the density profiles of the condensate in this case. The aspect ratio at $t=0$ in panel (a) and that at $t=3$ in panel (b) are similar. However, the vortex pair in Fig. \ref{fig:fig7}(b) has moved. The low-density region in the upper portion of the figure is due to a soliton through the propagation of a vortex pair.\cite{soliton} The motion of the vortex pair has little effect on the aspect ratio because the speed of expansion is larger than that of the pair. The velocity of the vortex pair $v$ is given by $v=\kappa/2\pi d$ with the distance $d$.
The velocity decreases with increasing value of $d$.
As the condensate expands, the distance $d$ increases to decelerate the vortex pair.
A critical distance $d_c$ of the vortex pair where $v$ is comparable to the velocity of the condensate is $d_c \sim 0.2$. Hence, the effect of the motion of the vortex pair can be negligible because the initial distance $d=4\xi=1.0$ is longer than $d_c$. 

It is important to compare our present work with the actual three-dimensional experiments by Henn {\it et al.}\cite{experiment} However, the comparison is not so easy because the cause of the SSE depends on the number of the vortices, the vortex configuration, the repulsive interaction, the distance of a vortex pair and so on. BECs prepared by Henn {\it et al.} are more elongated than our BECs. Thus, we actually prepared a BEC much elongated in order to investigate how many vortices the elongated BEC needs to reproduce the SSE. Although we put three vortex pairs along the minor axis of the condensate, the condensate split into a lot of fragments just after the expansion. This is because the size of the condensate is not much larger than the size of the vortices and the condensate with a large number of vortices cannot keep its own shape during the expansion. Hence, even when a condensate is elongated enough, a number of vortex pairs cannot be put along the minor axis to expand with keeping the shape of the condensate. This is a serious  difference between our two-dimensional simulation and the actual three-dimensional experiments. Therefore, we cannot obtain any definite conclusion on how many vortices are necessary for a actual three-dimensional BEC to make the SSE. 

\clearpage

\clearpage

\section{Conclusion}
In this paper, the expansion of a trapped two-dimensional BEC with vortices in non-turbulent state has been investigated qualitatively. Particular attention was focused to the SSE. It was confirmed that the aspect ratio of a vortex-free condensate inverts as a result of the anisotropy of the uncertainty principle along the $x$ and $y$ axes. Next, a single-vortex condensate was expanded to confirm that the inversion of its aspect ratio is faster than that of a vortex-free condensate. However, the single vortex does not introduce aother kind of anisotropy. If we prepare vorex pairs in the condensate, the aspect ratio is found to be increasingly suppressed with increasing numbers of vortex pairs. The anisotropy due to a vortex pair competes with that due to the uncertainty principle at $\lambda=0.87$, when the SSE is obtained.

Vortex pairs here create anisotropic superfluid flow that competes with the anisotropy due to the uncertainty principle, in order to create the SSE. Applying this understanding to a three-dimensional system, the SSE observed in the three-dimensional turbulent BEC\cite{experiment} can arise from some anisotropy of the vortex configuration. This anisotropy can come from two causes. First, anisotropic QT can be created by anisotropic vibration and rotation. Second, the QT is composed of a relatively small number of vortices leading to the large fluctuation of vortex configuration.

\end{document}